\documentclass{article}

\usepackage[super,sort&compress,comma]{natbib} 
\usepackage{mhchem}
\usepackage{times,mathptm}
\usepackage{sectsty}
\usepackage{balance}

\usepackage{graphicx} 
\usepackage{lastpage}
\usepackage{color}
\usepackage[format=plain,justification=raggedright,singlelinecheck=false,font=small,labelfont=bf,labelsep=space]{caption} 
\usepackage{fancyhdr}

\begin{document}

\thispagestyle{plain}
\renewcommand{\thefootnote}{\fnsymbol{footnote}}
\renewcommand\footnoterule{\vspace*{1pt}%
\hrule width 3.4in height 0.4pt \vspace*{5pt}} 
\setcounter{secnumdepth}{5}

\makeatletter 
\def\subsubsection{\@startsection{subsubsection}{3}{10pt}{-1.25ex plus -1ex minus -.1ex}{0ex plus 0ex}{\normalsize\bf}} 
\def\paragraph{\@startsection{paragraph}{4}{10pt}{-1.25ex plus -1ex minus -.1ex}{0ex plus 0ex}{\normalsize\textit}} 
\renewcommand\@biblabel[1]{#1}            
\renewcommand\@makefntext[1]%
{\noindent\makebox[0pt][r]{\@thefnmark\,}#1}
\makeatother 
\renewcommand{\figurename}{\small{Fig.}~}
\sectionfont{\large}
\subsectionfont{\normalsize} 

\renewcommand{\headrulewidth}{1pt} 
\renewcommand{\footrulewidth}{1pt}
\setlength{\arrayrulewidth}{1pt}
\setlength{\columnsep}{6.5mm}
\setlength\bibsep{1pt}

\noindent\LARGE{\textbf{Temperature Accelerated Monte Carlo (TAMC): a method for sampling the free energy surface of non-analytical collective variables.}}
\vspace{0.6cm}

\noindent\large{\textbf{Giovanni Ciccotti \textit{$^{a,b}$}
and Simone Meloni$^{\ast}$\textit{$^{d\ddag}$} }}\vspace{0.5cm}

\vspace{0.6cm}

\noindent \normalsize{We introduce a new method to simulate the physics of rare events. The method, an extension of the Temperature Accelerated Molecular Dynamics, comes in use when the collective variables introduced to characterize the rare events are either non-analytical or so complex that computing their derivative is not practical. 
We illustrate the functioning of the method by studying the homogeneous crystallization in a sample of Lennard-Jones particles. The process is studied by introducing a new collective variable that we call Effective Nucleus Size $\mathcal N$. We have computed the free energy barriers and the size of critical nucleus, which result in agreement with data available in literature. We have also performed simulations in the liquid domain of the phase diagram. We found a free energy curve monotonically growing with the nucleus size, consistent with the liquid domain. 
}
\vspace{0.5cm}

%

\section{Introduction}


\footnotetext{\textit{$^{a}$~Room 302B EMSC, School of Physics, University College Dublin, Belfield, Dublin 4, Ireland}}
\footnotetext{\textit{$^{b}$~Dipartimento di Fisica, Universit\'a ``La Sapienza',' Piazzale Aldo Moro 2, 00185 Rome, Italy}}
\footnotetext{\textit{$^{d}$~Room 302 EMSC, School of Physics, University College Dublin, Belfield, Dublin 4, Ireland. Tel. +353 (0)1 7161794}}
\footnotetext{\ddag~Permanent Address: Consorzio Interuniversitario per le Applicazioni di Supercalcolo Per Universit\`a e Ricerca (CASPUR), Via dei Tizii 6, 00185 Roma, Italy}
\footnotetext{\textit{$^{*}$~To whom the correspondence should be addressed: s.meloni@caspur.it}}

\label{sec:Introduction}
Over the last two decades several new methods have been introduced to sample the free energy surface as a function of a set of collective variables\cite{lerici1997, E-Li-VandenEijnden2004, Vanden-Eijnden:2009fk}. These methods have been applied to many challenging problems in chemistry,\cite{Park:2006gd, Lee:2006la} biology\cite{jacs-132_1010} and material science.\cite{monteferrante-2008} All these methods consist of an extended set of equations of motion coupling the dynamics of the atoms with that of a set of appropriate collective variables. In particular, the dynamics of the atoms is biased by a term that forces them to be in configurations compatible with the current realization of the collective variables. More explicitly, the coupling term is a function of the difference between the current value of the collective variable $\theta(x)$ and that of the corresponding additional dynamical variable $z$. Often the coupling is of the quadratic form $(k/2) (\theta(x) - z)^2$, where $k$ is the coupling parameter. The difference among these methods is in how the collective variables are forced to move out of metastable states. In the Temperature Accelerated Molecular Dynamics\cite{Maragliano2006168} (TAMD), of which the present method can be considered a direct extension, 
the variables associated with the collective variables are evolved at an artificially high temperature $\bar{T}$. Since the time required to overcome free energy barriers is roughly proportional to $\exp[\Delta F/k_B{\bar T}]$, where $\Delta F$ is the magnitude of barrier and $k_B$ is the Boltzmann constant, to higher $\bar T$ corresponds a shorter characteristic time to overcome them. By an adequate choice of $\bar T$ it is possible to make this time compatible with the maximum time achievable in atomistic simulations. A similar approach, but without the introduction of collective variables, has been investigated by VandeVondele and Rothlisberger\cite{cafes} and by Rosso et al. \cite{rosso:4389} 
In Metadynamics\cite{metadynamics, PhysRevLett.90.238302} the $z$ are forced to visit new states by biasing the free energy in the regions already visited by these auxiliary variables. 
 
In all these methods the atoms are evolved by molecular dynamics and this requires the calculation of the biasing force $-k(\theta(x)-z)\nabla_x\theta(x)$. This fact limits their application to problems described in terms of collective variables which are analytical with respect to atomic positions. In fact, in high dimension (a very typical case), the numerical calculation of the gradient $\nabla_x\theta(x)$  would be computationally really challenging. However, interesting cases of collective variables which are either non-analytical or for which the calculation $\nabla_x\theta(x)$ is either too complex or computationally too expensive do exist. An example of the first case is the ``ring-size'' collective variable, which is used in the study of formation of clathrates (gas hydrates) to distinguish this kind of crystals from ice \footnote[2]{In clathrates all the water molecules are part of four, five and six-member rings. In ice, no water molecule forms rings of any size.}\cite{clancy177} (which can also be formed in the same conditions). Examples of the second case are, in {\itshape ab initio} simulations, quantum mechanical observables which are not function of the Hamiltonian \footnote[4]{Be  $A(x)$ an observable defined as the expectation value of the operator $\mathcal{A}(r;x)$ over the electronic ground state $\psi(r;x)$ of the Hamiltonian $\mathcal{H}(r;x)$ ($ A(x) =  <\psi| \mathcal{A}(r;x) |\psi>$), where $r$ and $x$ are the electronic and atomic coordinates, respectively.  If the operator $\mathcal{A}(r;x)$ is a function of $r$ and $x$ via the $\mathcal{H}(r;x)$ ($\mathcal{A}(r;x) = \mathcal{A}(\mathcal{H}(r; x))$), then, following the Hellman and Feynman theorem, ${d A(x) \over d x} = <\psi| {d\mathcal{A}(r;x) \over d \mathcal{H}(r;x)} {\partial \mathcal{H}(r;x) \over \partial x} |\psi>$. However, if $\mathcal{A}(r;x)$ does not depend on $x$ via a function of $\mathcal{H}(r;x)$, ${d A(x) \over d x}$ depends also on the derivative of the ground state wavefunction with respect to $x$: ${d A(x) \over d x} = <\psi| {\partial \mathcal{A}(r; x) \over \partial x} |\psi> + <\psi| \mathcal{A}(r;x) |{\partial \psi \over \partial x}> + <{\partial \psi \over \partial x}| \mathcal{A}(r;x) |\psi>$. This implies that if $\mathcal{A}(r;x)$ is not a function of $\mathcal{H}(r;x)$, the use of the observable $A(x)$ as a collective variable in the biased dynamics will require the calculation of the perturbed wavefunction $\partial \psi(r;x) \over \partial x$.
}. In fact, in this case some perturbation theory method should be applied to calculate the biasing force.\cite{PhysRevA.52.1096}

The aim of this paper is to introduce an extension of TAMD, let us call it Temperature Accelerated Monte Carlo (TAMC), which allows to treat these more general cases. Since TAMC is inspired by the TAMD, and is based on the same assumptions, we first revise TAMD (Sec. \ref{sec:TAMD}), then introduce TAMC (Sec. \ref{sec:TAMC}), and finally illustrate the method by showing how TAMC allows to study homogeneous nucleation as described by a new collective variable, introduced in this paper, which we call Effective Nucleus Size (ENS) (Sec. \ref{sec:Nucleation}).

\section{TAMD: Temperature Accelerated Molecular Dynamics method
\label{sec:TAMD}
}
In TAMD we introduce the following set of coupled equations (for simplicity, we denote $x$ and $z$ as scalar variables but are, indeed, vectors of suitable dimension):
\begin{eqnarray}
\label{eq:TAMD}
m {\ddot x} &=& -\nabla_x U_k(x,z) + thermo(\beta) \nonumber \\
\mu {\ddot z} &=& -\nabla_z U_k(x,z) + thermo({\bar \beta})
\end{eqnarray} 

\noindent
where $m$ is the physical mass, $\beta = \left (k_B T\right )^{-1}$ and $thermo(\beta)$ indicates that the atoms are coupled to a thermostat at $\beta$. $\mu$ is the inertia of $z$, a parameter that can be tuned so as to achieve the adiabatic separation of the $x$ dynamics with respect to that of $z$ (see below), ${\bar \beta} = \left (k_B {\bar T}\right )^{-1}$ and $thermo({\bar \beta})$ indicates that the auxiliary variables $z$ are coupled to a thermostat at $\bar \beta$. The potential $U_k(x,z) = V(x)  + k/2(\theta(x)-z)^2 $ is the sum of the physical and a biasing potential. In the limit in which $z$ is much slower than $x$, the force acting on $z$ ($-\nabla_z U_k(x,z) = k(\theta(x(t)) - z) = f_k(z)$) can be substituted by the time-averaged force:

\begin{equation}
\label{eq:effectiveForce}
{\bar f}_k(z) = \lim_{\tau \rightarrow \infty}{1 \over \tau} \int_0^\tau dt k(\theta(x(t)) - z)   =  {\mathcal Z}^{-1}_k(z) \int dx k(\theta(x) - z) \exp[-\beta U_k(x,z)] 
\end{equation}

\noindent where ${\mathcal Z}_k(z) = \int dx \exp[-\beta U_k(x,z)]$. The inertia $\mu$ can be tuned so as to obtain an adequate separation of the characteristic evolution times of $z$ and $x$. The second equality in eq. \ref{eq:effectiveForce} stems from the assumption that, apart for the $z$, the remaining degrees of freedom of the system are ergodic. The effective force can be interpreted as the derivative of the effective potential $F_k(z) = -\beta^{-1} ln[{\mathcal Z}_k(z)/\mathcal{Z}]$, where ${\mathcal Z} = \int dx \exp[-\beta V(x)]$ is the canonical partition function of the real system. Since $\mathcal Z$ is $z$-independent its introduction does not affect our argument but it is necessary for the interpretation of the effective potential $F_k(z)$ as a free energy. Noting that $\lim_{k\rightarrow \infty }\exp[-\beta {k \over 2} (\theta(x) - z)^2 ] / \sqrt{2\pi / \beta k}= \delta(\theta(x) - z)$, 
in the limit of $k \rightarrow \infty$ $F_k(z) \rightarrow -\beta^{-1} ln[P_{\theta}(z)] = F(z)$, where $P_{\theta}(z)$ is the probability density function that the system is in the state $\theta(x) = z$. In other words, for $k$ sufficiently large, $z$ is a set of random variables moving on the free energy surface and it is therefore distributed according to 

\begin{equation}
\label{eq:TAMDPDF}
P_{\theta}(z) = \exp[-{\bar \beta} F(z)]
\end{equation}

\noindent Since the time average in eq. \ref{eq:effectiveForce} is taken over the dynamics os the $x$'s thermalized at $\beta$, the $z$ is distributed according to the free energy $F(z)$ at the physical temperature. However, since in eq. \ref{eq:TAMDPDF} the free energy $F(z)$ is multiplied by $\bar \beta << \beta$, the sampling of the unlikely regions is enhanced and the $z$ can quickly overcomes the barriers on the physical free energy surface $F(z)$.

\section{TAMC: Temperature Accelerated Monte Carlo method
\label{sec:TAMC}
}
To introduce TAMC we start by observing that the key point in TAMD is that the (slow) variable $z$, being driven by  $k (\theta(x) - z)$, evolves indeed according to the effective force defined in the r.h.s. of eq. \ref{eq:effectiveForce}, which is the average of $k (\theta(x) - z)$ over the canonical ensemble for the biased potential $U_k(x,z)$. In TAMD this ensemble average is computed by (adiabatically separated) molecular dynamics (MD). However, this ingredient is not crucial and the MD on the $x$'s could be replaced by any adequate sampling technique. For example, we could replace the $x$'s MD by Monte Carlo (MC). In TAMC we take advantage of this freedom to replace MD by MC, which does not require the calculation of $\nabla_x \theta(x)$ for the evolution of the $x$ and can therefore be used in combination with  non-analytical collective variables. Let us imagine to evolve the dynamics of the collective variables according to the Langevin dynamics, then the TAMC can be expressed as follows

\begin{equation}
\label{eq:TAMC}
\mu \ddot z =  k (\theta(x_{MC}) - z) - \gamma \dot z + \sqrt{2{\bar \beta} \gamma} \eta(t)
\end{equation}

\noindent where $\gamma$ is the friction coefficient and $\eta(t)$ is a Gaussian process with mean 0 and covariance $\left <\eta(t)\eta(t')\right> = \delta(t-t')$. In eq. \ref{eq:TAMC} the subscript $MC$ on the $x$ indicates that the force $k (\theta(x_{MC}) - z)$ is computed according to the atomic configuration evolved by MC, in parallel to $z$, according to the potential $U_k(x,z)$. If $z$ is slow with respect to $x$, in a sense that will be made precise below, $z$ will evolve according to the effective force $<k (\theta(x) - z)>_{U_k}$. Following the same argument of Maragliano and Vanden-Eijnden,\cite{Maragliano2006168} in the limit $k \rightarrow \infty$ $z$ is distributed according to $\rho(z) = \exp[-{\bar \beta}F(z)]$. It remains to make unambiguous the concept of adiabatic separation in TAMC. In TAMD, where both $x$ and $z$ follow a proper dynamics, adiabatic separation means that the characteristic time of $z$ ($\tau_z$), i.e. the time required for a significant displacement of $z$, is much longer than the characteristic time of $x$ ($\tau_x$): $\tau_z >> \tau_x$. In TAMC $x$ is evolved by MC and therefore the definition of adiabatic separation mentioned before cannot be applied. We need to introduce a different definition. 
Let $h$ be the timestep used for the numerical integration of eq. \ref{eq:TAMC}, then $n_z = \tau_z / h$ is the number of timesteps required for a significant displacement of $z$. let $n_x$ be the number of MC steps required for a accurate sampling of the probability $\rho(x|z) = \exp[-\beta U_k(x,z)]/{\mathcal Z}_k(z)$ at a given z.  Then, if $n_z >> n_x$, $z$ evolves according to the effective force $<k (\theta(x) - z)>_{U_k}$. We stress again that, at variance with TAMD (or, for that matter, Metadynamics or Adiabatic Dynamics), the evolution of the atomistic configuration in TAMC does not require the gradient of the collective variable $\nabla_x\theta(x)$. This method, therefore, applies also to non-analytical collective variables.

\begin{figure}[h]
\begin{center}
\includegraphics[width=0.45\textwidth]{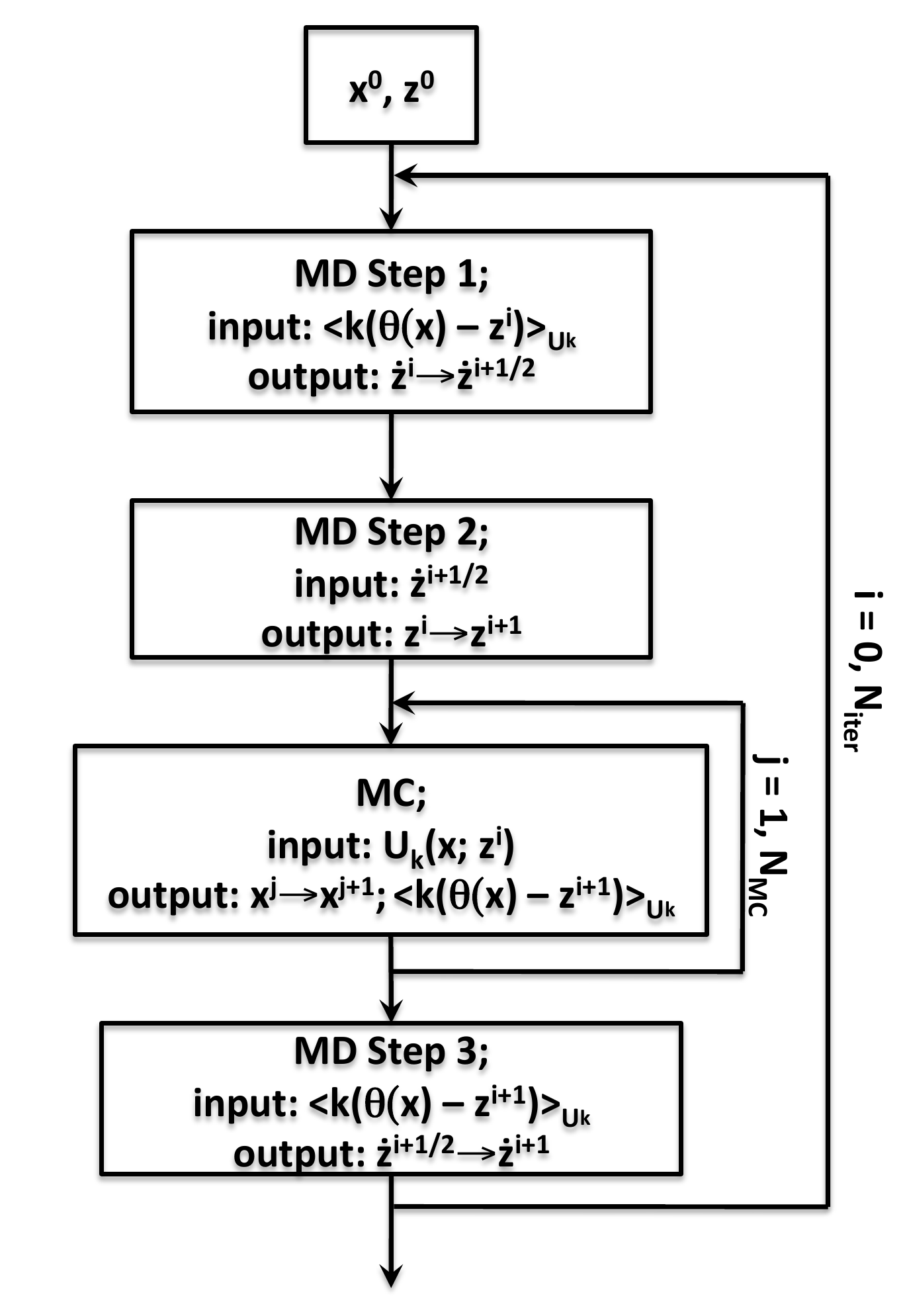}
\caption{Flowchart of the TAMC algorithm. In this chart we assume that the MD step is performed by a three-step algorithm, in particular we used the Vanden-Eijnden and Ciccotti algorithm for integrating Langevin equation of motion.\cite{langevinIntegrators}}
\label{Fig:TAMCAlgorithm}
\end{center}
\end{figure}

{The algorithm implementing the method described above is quite simple and, in practice, consist in a dynamics over the collective variables $z$ with the force $k (\theta(x_{MC}) - z)$ computed according to the atomic configuration evolved by a Metropolis Monte Carlo controlled by the potential $U_k(x,z)$. This algorithm is presented schematically in Fig. \ref{Fig:TAMCAlgorithm}. Starting from the atomic configuration $x^i$ and the value $z^i$ of the collective variable and its velocity ${\dot z}^i$, the values $\dot z^{i+1/2}$ at the ``half time step'' and $z^i$ at the next time step are computed according to the force $k (\theta(x^i) - z^i)$. After this, the atomic position is evolved according to the Metropolis MC scheme: i) an atom is chosen at random, ii) a random displacement is applied on this atom so obtaining the configuration $x^*$, iii) the variation in the (biased) potential energy $\Delta U_k = U_k(x^*,z^{i+1}) - U_k(x^i,z^{i+1})$ is computed, iv) the move is accepter (rejected) according to the usual Metropolis criterion: either with probability 1 if the $\Delta U_k < 0$ or with a probability equal to $\exp[-\beta \Delta U_k]$. The only difference with respect to the standard MC move is that in TAMC the variation of energy is the sum of the variation of the interatomic potential, which is computed according to the standard procedure, plus the sum of the biasing term $k/2(\theta(x) - z)^2$. This procedure is repeated for $N_{MC}$ steps and the force $<k (\theta(x) - z)>_{U_k}$ is computed as the average of $k (\theta(x) - z)$ over the configurations generated by the MC sampling. We would like to remark that any method developed for standard MC can also be applied to TAMC. In particular, we can compute not only the Helmhotz free energy but also the Gibbs free energy by running constant pressure MC simulation, as it was done in this paper to study the crystal nucleation (see Sec. \ref{resultsAndDiscussion}).}   

We now compare TAMD and TAMC from the point of view of the various parameters governing the simulation. In TAMD the timestep $h$ is fixed by the timescale of the fast degrees of freedom, that is the atoms. Typically, the timestep is of the order of $10^{-15}~s$ ($1$~fs). The parameter $\mu$, the inertia of the variable $z$, is adjusted so as to achieve the adiabatic separation of the dynamics of the $z$ and the $x$. Depending on the type of thermostat, other parameters must be adjusted so that atoms are adiabatically at the thermal equilibrium. For example, if eq.s \ref{eq:TAMD} take the form of Langevin equations, the friction parameters $\gamma$ and $\gamma'$, for the $x$ and the $z$, need to be adjusted. If eq.s \ref{eq:TAMD} take the form of Nos\'e-Hoover equations the inertia $Q$ and $Q'$ of the thermostat variables need to be defined. These parameters control the timescale over which in eq. \ref{eq:effectiveForce} we can assume the time average to be equivalent to the ensemble average.
In TAMC we must chose the MD and MC parameters controlling the evolution of $z$ and $x$, respectively, such that the adiabatic separation as described before (i.e. $n_z >> n_x$) is achieved. The MC on the $x$ is governed by the usual parameters (maximum atomic displacement for standard MC) according to the usual criteria on the acceptance ratio (typically between 30 and 60~\%). The choice of these parameters determine the number of MC steps $n_x$ needed between the Langevin timesteps of the $z$. Once $n_x$ is determined, the inertia $\mu$, thermostat parameters (e.g. $\gamma$ for Langevin and $Q$ for Nos\'e-Hoover) and the length of the MD timestep are set such that the adiabatic separation is achieved.

\section{An example of application of TAMC: homogeneous nucleation in supercooled Lennard-Jones liquids. 
\label{sec:Nucleation}}

In this section we shall illustrate how TAMC works in practice by studying the homogeneous nucleation in a Lennard-Jones liquids. The nucleation is studied as a function of one collective variable monitoring the size of the crystalline nucleus (see Sec. \ref{sec:NucleationCollCoord}). Since we use only one collective variable, other method could be used as well to tackle this problem. Indeed, the nucleation in Lennard-Jones liquids, using different collective variables, has been already studied using the Umbrella sampling and the Partial Path Transition Interface Sampling techniques (see Refs. [\citetext{15}] and [\citetext{17}]).

\subsection{Collective variable for nucleation}
\label{sec:NucleationCollCoord}
One collective variable proposed to study homogeneous nucleation of liquids is the size ${\mathcal N}(r_1, \dots, r_N)$ of the largest cluster of solid-like particles\cite{wolde:9932,lechner:114707, PhysRevLett.94.235703} expressed in terms of number of particles. 
In the present work we found more suitable to adopt an improved definition of the size of a crystalline nucleus. Our definition, and its advantages over that used in the referred works, are given in the following. 
As in previous works \cite{wolde:9932, PhysRevLett.94.235703}, to define ${\mathcal N}(r_1, \dots, r_N)$ we start from the local bond-orientational order parameter of Steinhardt et al.\cite{PhysRevB.28.784}

\begin{equation}
\label{eq:locaBondOrder}
q_{lm}(i) = \sum_{j=1}^{N_i} Y_{lm}({\vec r}_{ij})
\end{equation}

\noindent where $Y_{lm}({\vec r}_{ij})$ is the spherical harmonics of order $l$ and $m$ computed at the polar and azimuthal angles associated to the vector ${\vec r}_{ij}$ connecting atoms $i$ and $j$. The sum runs over the $N_i$ nearest neighbors of the atom $i$. Here and in the following, the nearest neighbors of an atom, say $i$, are those atoms satisfying the condition $|{\vec r}_{ij}| \leq 1.5~\sigma$. 
For a proper choice of $l$ \footnote[5]{A first aspect that must be taken into account in the choice of $l$ is that spherical harmonics corresponding to odd $l$ are anti-symmetric under inversion ($Y_{lm}({\vec r}) = -Y_{lm}(-{\vec r}) $). In many crystals the nearest neighbors of an atom $i$ have an equilibrium configuration such that if one of them is connected to $i$ by the vector $\vec r_{ij}$, there is another one connected by $-\vec r_{ij}$ (all the simple lattices - simple cubic, FCC, BCC, etc - have this property). This implies that the $q_{lm}$'s with odd $l$ of these systems are all $0$. Then, for $q_{lm}$ to be useful in distinguishing between different local environments, $l$ must be even. Steinhardt et al.\cite{PhysRevB.28.784} for bulk and cluster systems at their equilibrium configuration and ten Wolde et al.\cite{wolde:9932} for nucleation have shown that the optimal $l$ to use for $q_{lm}$ and derived observables (introduced to distinguish between systems of different symmetry) are $l=4$ or $l=6$. In this paper we use $l=6$ to identify connected particles, as this is consistent with the $l$ used in Ref.   
[\citetext{17}]. That choice 
allows a direct comparison with previous results.}, $q_{lm}(i)$ characterizes the local structure around the atom $i$. In crystals the $q_{lm}$ of neighboring atoms are similar (coherent). In liquids the degree of coherence, in a sense made precise below, is much lower than in crystals. We will use this empirical observation to distinguish between crystal-like from liquid-like particles. The degree of coherence of $q_{lm}$ of neighboring atoms can be measured by
 
\begin{equation}
\label{eq:correlation}
C_{ij} = {\left |  \sum_{m=-l}^{l} q_{lm}(i)^* q_{lm}(j)  \right | \over \left | q_l(i) \right | \left | q_l(j) \right |  }
\end{equation}

\noindent where  $\left | q_l(i) \right |$ and $\left | q_l(j) \right |$ are defined as $\left | q_l(i) \right | = \sqrt{ \sum_{m=-l}^{l} q_{lm}(i)^* q_{lm}(i)}$. {$q_{lm}(i)^*$ is the complex conjugate of $q_{lm}(i)$}. In fact, $C_{ij}$ can be interpreted as the normalized dot product between the vectors $q_{l}(i)$ and $q_{l}(j)$, with components $q_{lm}(j)$. Therefore $C_{ij}$  measures how much the two vectors are parallel. When two atoms have the same environment their $q_{lm}$ are the same and $C_{ij}=1$. This is the case of two neighboring atoms in a crystal at $0$~K. On the contrary, when two atoms have a different environment, as it is typically the case in disordered systems, $C_{ij}$ is much smaller. At finite temperature  $C_{ij} \neq 1$ also in crystals. However, it has been empirically observed that $C_{ij}$ is $\geq 0.5$ in crystalline samples at finite temperature.\cite{wolde:9932} According to this, we say that two particles with a $C_{ij}$ above this threshold are ``connected''. We shall use throughout the paper the word ``connected'' in this strict sense. Neighboring particles with an alike environment ($C_{ij} \geq 0.5$) are found also in liquids, but in crystals the number of connected particles around a given atom (typically $\geq 7$) is higher than in liquids (typically below $4 - 5$). So, we can use the number of connected particles as a parameter to distinguish between liquid-like and crystal-like particles. 

\begin{figure}[h]
\begin{center}
\includegraphics[width=0.45\textwidth]{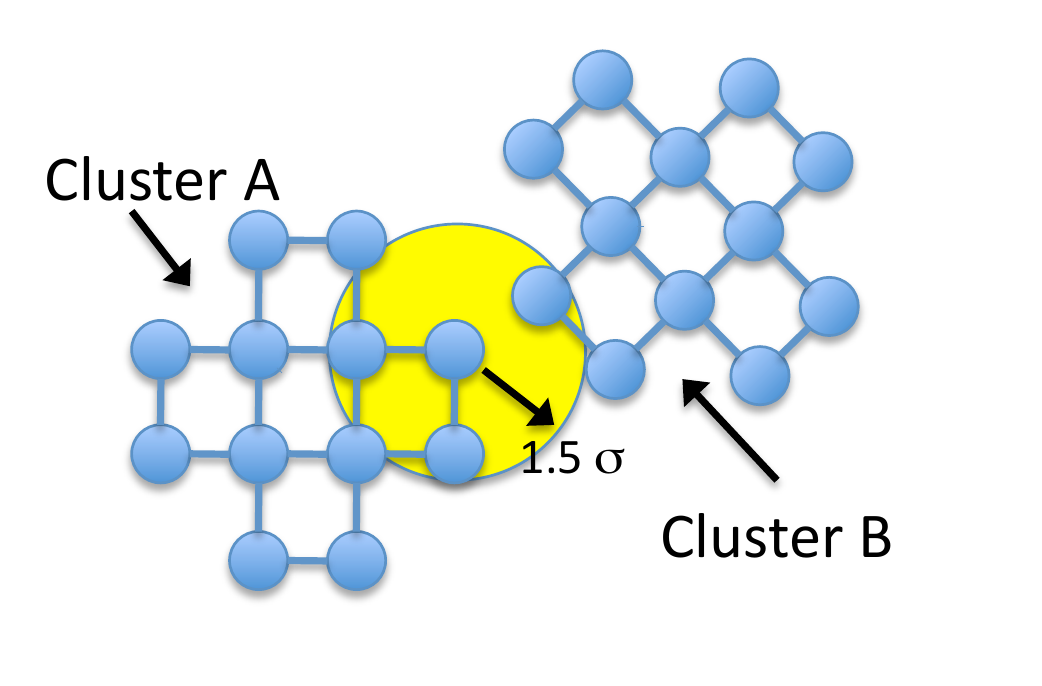}
\caption{ Schematic representation of atoms forming clusters and their connectivity. The clusters `A' and `B' are two separated clusters as no one atom in one cluster is connected with any atom in the other, even though there are atoms of one cluster which are nearest neighbors of atoms of the other cluster. This is because the orientation of atoms in one cluster are different from the one in the other and upon growing they will not form a single crystal.}
\label{fig:clusters}
\end{center}
\end{figure}

Once we have identified crystal-like particles, we can search for clusters of such particles in the sample. 
In Refs. [\citetext{15}] and [\citetext{17}] 
a crystal-like particle constitute, or is assigned to, a crystal nucleus if it is within a given distance ($1.5~\sigma$) from another crystal-like particle (possibly of an already existing nucleus), whether this last particle is connected or not to it. On this, our definition of crystal nuclei is different. We define a nucleus as the set of crystal-like particles which are connected (and therefore close, in the sense defined above) with at least another particle in the cluster (see Fig. \ref{fig:clusters}). According to the definition given in 
Refs. [\citetext{15}] and [\citetext{17}] 
the nucleus might include also particles which are not connected, i.e. particles which have a different environment. On the contrary, our definition of clusters allows to distinguish between particles which are indeed members of the same nucleus from particles which might be member of neighboring nuclei. The two different nuclei, if growing, will eventually form a grain boundary, like the nuclei `A' and `B' of Fig. \ref{fig:clusters}. {We expect that by adopting our definition of crystal nucleus the size of critical nuclei observed in Ref. [\citetext{15}] and [\citetext{17}]  would be reduced in size by $\sim 10-50$ atoms, which is the typical size of subcritical crystal nuclei usually found in Lennard-Lones liquids (see top panel of Fig. \ref{fig:ClustersizeTimeline}) and that could be close in space to the largest connected crystalline cluster. Indeed, the fact that the definition of Ref. [\citetext{17}] might include in the crystalline nucleus also particles that are not connected to it could account for the broad shape of the committor observed in this paper. In fact, as explained before, in this case clusters of significantly different sizes, both under and super-critical, could be assigned to the critical nucleus.}
In practice, we identify these clusters by using methods of the Graph theory. In our approach the crystal-like particles are the nodes and the connections between them are the edges of a graph. We then identify clusters in the graph by means of the Deep First Search method.\cite{DE-Knuth} The Deep First Search method consists in searching the graph for connected nodes starting from a root node, and exploring as far as possible along the edges of the graph. Once all the particles directly or indirectly connected to the root are identified the search for the members of the present cluster is completed. The search for the particle belonging to another cluster is then started by defining a new root among the crystal-like particles not yet assigned to any nucleus. This search is repeated until all the crystal-like particles have been assigned to a cluster. 
The collective variable ${{\mathcal N}}(r_1, \dots, r_N)$ could be defined  as the number of particles forming the largest cluster. It appears clear from the description above that this collective variable is non-analytical as it does depend on atomic positions $r$ but not through an explicit formula, rather through a search procedure. Indeed both definition of nucleus size, the present one and the one given in  
Refs. [\citetext{15}] and [\citetext{17}], 
are non-analytical.

\begin{figure}[h]
\begin{center}
\includegraphics[width=0.3\textwidth]{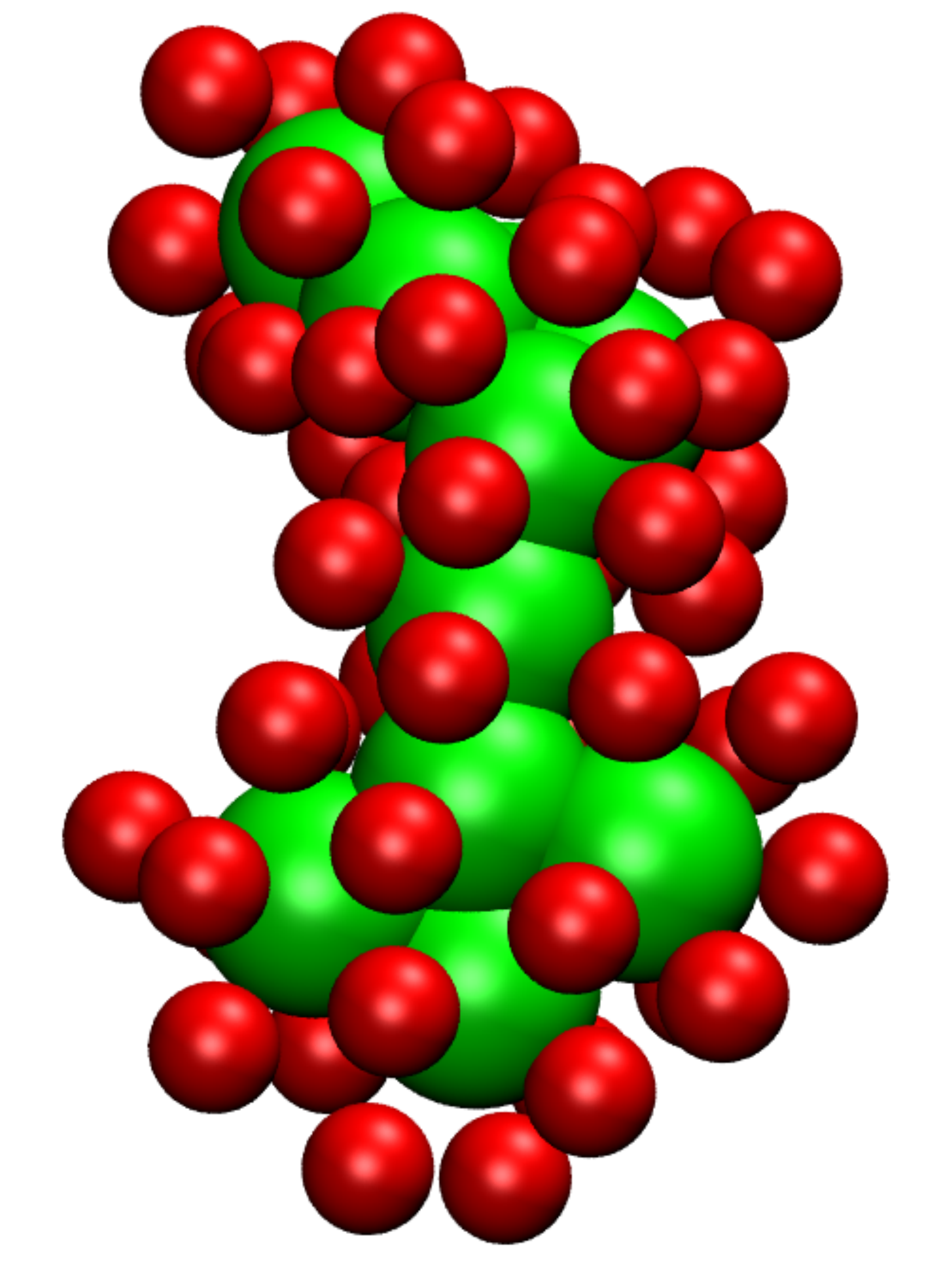}
\caption{ Nucleus and buffer particles in a sub-critical nucleus. For sake of readability, nucleus particles are bigger (and green online) and buffer particle are smaller (and red online). The sum of eq. \ref{eq:DefN} runs over both sets of atoms.}
\label{fig:cluster+buffer}
\end{center}
\end{figure}

The definition of the collective variable ${{\mathcal N}}(r_1, \dots, r_N)$ given above is not very efficient to use with TAMC since it is discrete and therefore the associated potential $(k/2) ({{\mathcal N}}(r_1, \dots, r_N) - z)^2$ biases the simulation only when the MC move changes its value. When this does not occur, the MC is unbiased and there is no acceleration on the sampling. Let us illustrate this problem by an example. Let us assume that ${{\mathcal N}}(r_1, \dots, r_N)$ is lower than $z$. This means that a MC move increasing ${{\mathcal N}}(r_1, \dots, r_N)$ should be favored by the biasing potential. An MC move increases ${{\mathcal N}}(r_1, \dots, r_N)$ if it connects a crystal-like particle not yet member of the nucleus to another particle which is in the nucleus or if it transforms a non crystal-like particle connected to the nucleus into a crystal-like one. However, both processes might require several steps. For example, the first might occur through a series of MC moves which gradually increase the degree of crystallinity of a non crystal-like particle (the meaning of degree of crystallinity, made precise below, for the time being must be understood as the number of particles connected to the present one). However, as far as this particle does not become crystal-like, all these moves are unbiased, which slows down the convergence of the MC procedure. Moreover, since the definition of connectivity and crystallinity of a particle is somewhat arbitrary, as it depends on the value of the lower bound of $C_{ij}$ for which we consider two particles connected, the current value of the nucleus size is strongly dependent on this arbitrary choice. We solved the problem of biasing all the MC moves and alleviated the problem of the arbitrariness of ${{\mathcal N}}(r_1, \dots, r_N)$ by introducing a continuous version of the collective variable described above. This continuous version of the collective variable we have called Effective Nucleus Size. It is defined by the following equation

\begin{equation}
\label{eq:DefN}
{\mathcal N}(r_1, \dots, r_N) = \sum_{i \in cluster + buffer} w_c(i) w_n(i) 
\end{equation} 

\noindent where the sum runs over the members of the largest cluster as defined above plus the atoms laying within a $1.5~\sigma$ thick buffer around the cluster (see Fig. \ref{fig:cluster+buffer})
and $w_c(i)$ and $w_n(i)$ are weights accounting for the degree of connection of the particle $i$ to the cluster and its degree of crystallinity, respectively. The buffer particles are those particles which do not belong to the nucleus and which are within a distance of $1.5~\sigma$ from any nucleus particle. The so defined ${\mathcal N}(r_1, \dots, r_N)$ is still non-analytical as the sum in eq. (\ref{eq:DefN}) can be defined only after the atoms in the largest cluster are identified through the procedure explained before. 

\begin{figure}[h]
\begin{center}
\includegraphics[width=0.45\textwidth]{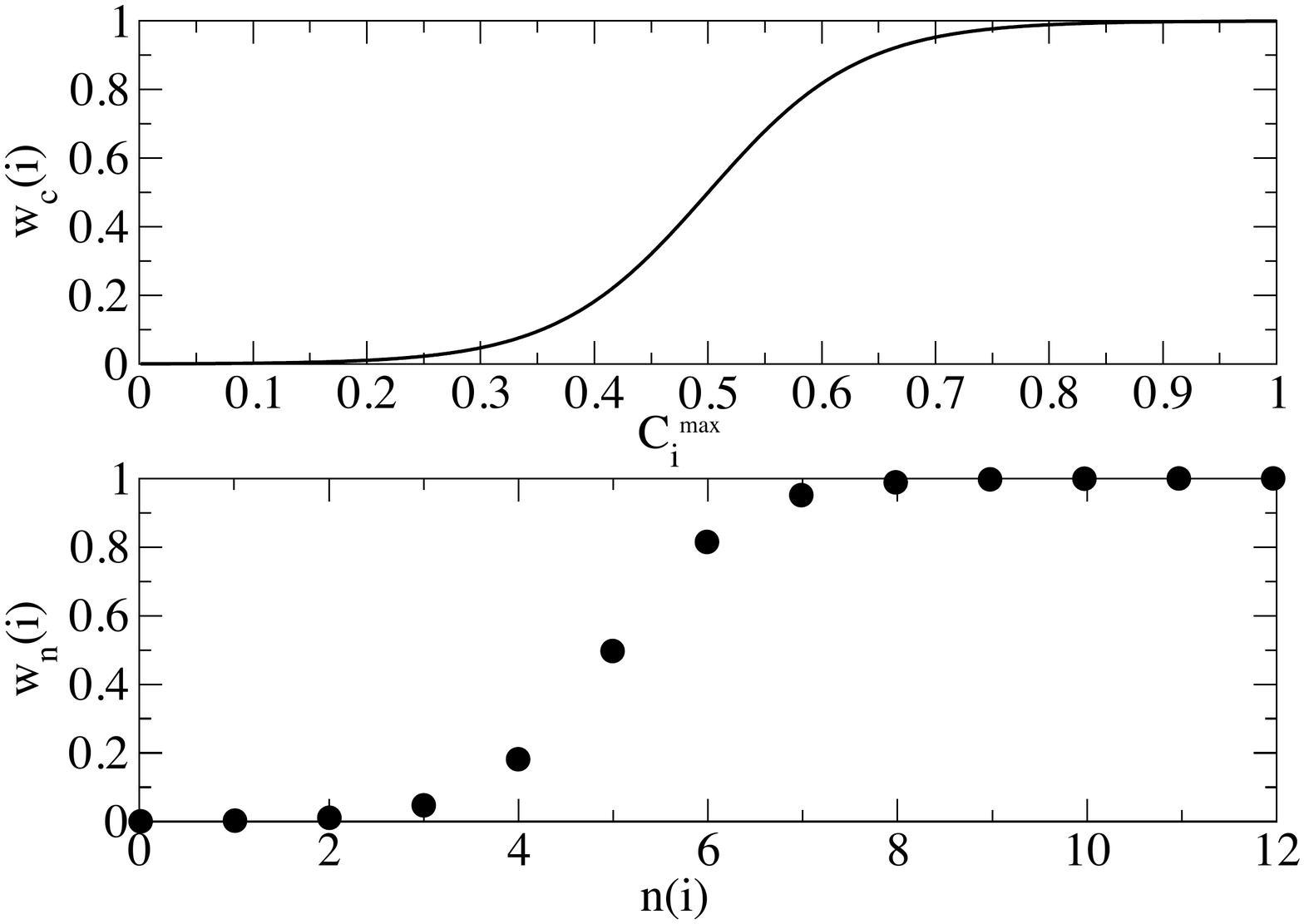}
\caption{$w_c$ and $w_n$ weights as a function of the degree of connection $C_{i}^{\max}$ and the degree of crystallinity $n$ (see text).}
\label{fig:weights}
\end{center}
\end{figure}

In order to give an explicit expression for $w_c(i)$ we need to analyze what are the expected properties of this parameter. $w_c(i)$ must tend to $0$ (to $1$) when the particle $i$ is loosely (strongly) connected to the cluster, and must smoothly go from one limit to the other in between. In other words, it should be a sigmoid with respect to the degree of connection of the particle $i$ to the cluster. The parameter accounting for the degree of connection in our modeling of nucleation is $C_{i}^{\max} = \max_j(C_{ij})$, where $j$ is a particle belonging to the cluster.  A reasonable function $w_c(i)$ of $C_{i}^{\max}$ with the properties mentioned above can be obtained from the Fermi function

\begin{equation}
w_c(i) = 1 - {1 \over \exp[\lambda_c (C_{i}^{\max} - \mu_c)] + 1}
\end{equation}

\noindent where $\mu_c$ is the parameter controlling the value of $C_{i}^{\max}$ at which the function switches from low values to high values and $\lambda_c$ is a parameter controlling the smoothness of the switching. We set  $\mu_c = 0.5$.
$\lambda_c$ was set to $15$ so that $w_c(i)$ is very low ($\leq 0.05$) for $C_{i}^{\max} \leq 0.3$ and   close to 1 ($\geq 0.95$) for $C_{i}^{\max} \geq 0.7$ (see the top panel of Fig. \ref{fig:weights}). 
Similarly, $w_n(i)$ is defined according to the following expression

\begin{equation}
w_n(i) = 1 - {1 \over \exp[\lambda_n (n(i) - \mu_n)] + 1}
\end{equation}

where $n(i)$ is the number of neighbors connected to the particle $i$. $n(i)$, as announced above, measures the degree of crystallinity of a particle. The $\mu_n$ and $\lambda_n$ are set to $5$ and 1.5, respectively. With this choice, $w_n(i)$ is close to $1$ (0.95) for $n(i) = 7$ (see the bottom panel of Fig. \ref{fig:weights}), which is consistent with the lower bound used to identify crystal-like particles for the identification of nucleus.

The continuous formulation of the collective variable ${\mathcal N}(r_1, \dots, r_N)$ given above solves the problems of the original definition. In fact, with this new definition, essentially any MC move is biased. In order to illustrate this, let us imagine that the current value of ${\mathcal N}(r_1, \dots, r_N)$ is lower that the target value $z$. Any  MC move involving atoms belonging to the largest cluster or to the buffer region (see Fig. \ref{fig:cluster+buffer}) which increases the connection of this particle with the cluster or increase its degree of crystallinity will increases ${\mathcal N}(r_1, \dots, r_N)$ and therefore will be favored as they reduce the biasing potential, even though the moved particle is not yet crystal-like according to the definition given before or it is not yet connected to the nucleus. On the contrary, any MC move reducing the connection of the particle to the cluster or its degree of crystallinity will be disfavored. In other words, since the change of the collective variable is continuous (indeed almost continuous, since it is continuous for the term depending on the connectivity but not for the one depending on the number of neighbors, which is function of the discrete variable $n(i)$) the possibility of accepting MC moves that bring the system closer to the target value is higher than with the original discrete formulation of the collective variable ${{\mathcal N}}(r_1, \dots, r_N)$. This definitely improves the efficiency of the sampling.

\subsection{Results and Discussion
\label{resultsAndDiscussion}
}

\begin{figure}[h]
\begin{center}
\includegraphics[width=0.45\textwidth]{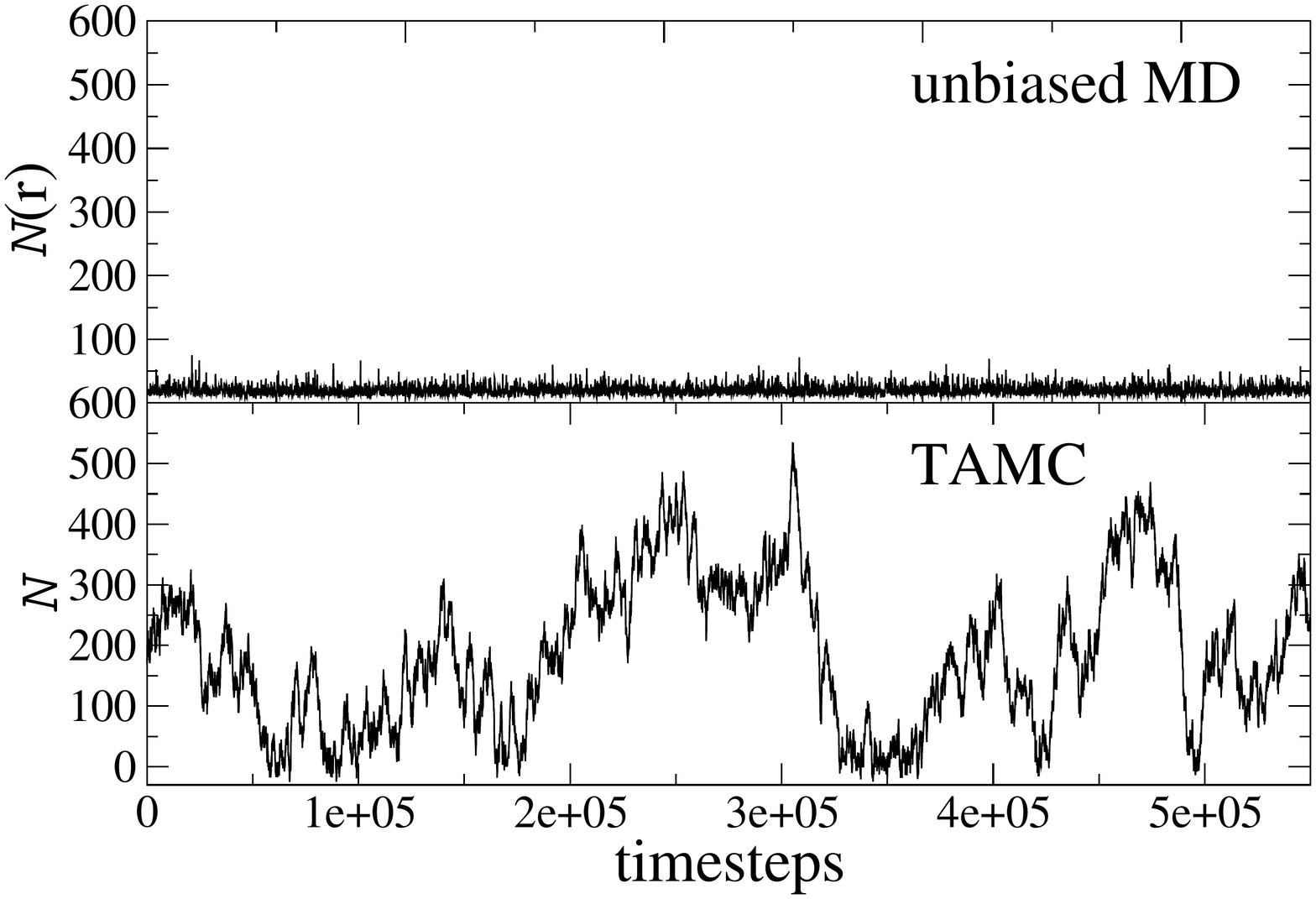}
\caption{Timeline of the collective variable ${\mathcal N}(r_1, \dots, r_N)$ obtained from standard MD simulations (top panel) and by TAMC (bottom panel).}
\label{fig:ClustersizeTimeline}
\end{center}
\end{figure}

We ran unbiased and TAMC simulations of a liquid sample of 3456 Lennard-Jones particles. The pressure and the temperature were kept fix at $P = 5.6$ $T = 0.92$, i.e. we performed isobaric-isothermal MD/MC simulations  (pressure and temperature are reported in Lennard-Jones units). These conditions correspond to a 17\% degree of supercooling, i.e. at this pressure the temperature is 17\% lower than the melting temperature (we used the Hansen and Verlet\cite{PhysRev.184.151} data to estimate the melting temperature). This pressure and temperature are in the range used in other computational investigations of nucleation in Lennard-Jones systems.\cite{wolde:9932, PhysRevLett.94.235703,lechner:114707} The liquid sample is obtained by melting a Body Centered Cubic crystal at high temperature ($T=50$) and then cooling it down to T=0.92 very slowly.

We performed unbiased MC and MD simulations of the liquid and monitored the ${\mathcal N}(r_1, \dots, r_N)$ (see the top panel of Fig. \ref{fig:ClustersizeTimeline}, where only the MD results are shown). In these simulations, on a trajectory of 55000 MD timesteps (or $2\times10^6$ steps for MC) we only observed the ${\mathcal N}(r_1, \dots, r_N)$ to fluctuate around 20. Nuclei of this size are under-critical, that is their size is smaller than the size corresponding to the maximum of the free energy versus $\mathcal N$ curve, which Moroni et al.\cite{PhysRevLett.94.235703}, using their definition of ${\mathcal N}$, have found to be $\sim 250$ for a Lennard-Jones liquid in similar conditions (25\% degree of supercooling). This result confirms that nucleation is a rare event and that relevant information on this process, such as the free energy barrier and the critical size of the nuclei, cannot be obtained by brute force simulations in the conditions of ``moderate'' supercooling. 

\begin{figure}[h]
\begin{center}
\includegraphics[width=0.45\textwidth]{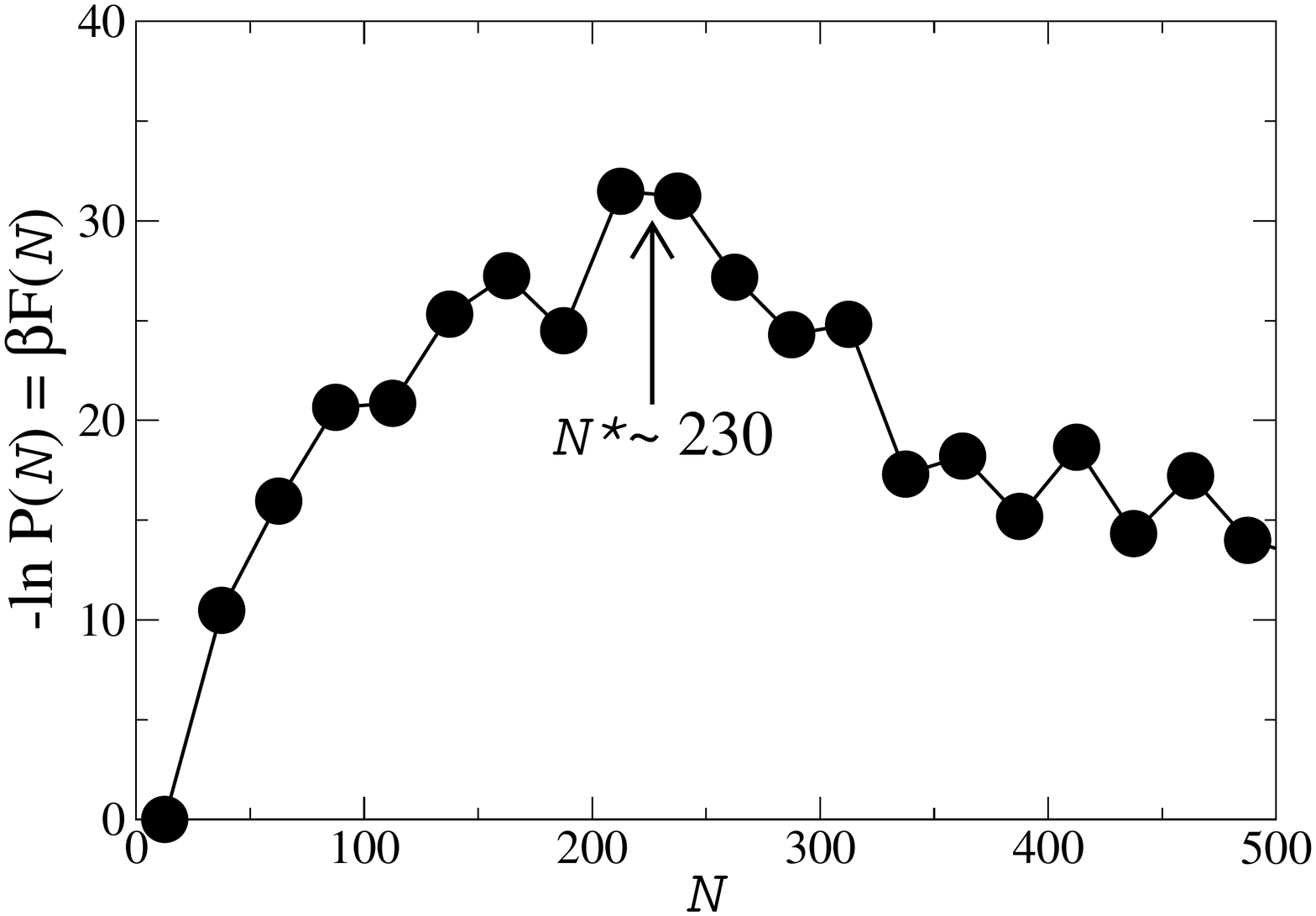}
\caption{Free energy vs. $\mathcal N$ curve at P=5.6 and T=0.92. The nucleus critical size is also reported.}
\label{fig:FvsN}
\end{center}
\end{figure}

Starting from the same liquid sample, we run a TAMC simulation using  ${\mathcal N}(r_1, \dots, r_N)$ as collective variable. In these simulations we evolved the dynamical variable ${\mathcal N}$ associated to ${\mathcal N}(r_1, \dots, r_N)$ (see Eq. \ref{eq:TAMC}) every 3456 MC steps on the nuclei. This roughly corresponds to make one ${\mathcal N}$ move every one move of all the atom. As a first remark, we notice that with TAMC we are able to explore a wide range of the collective variable space, as shown in the bottom panel of Fig. \ref{fig:ClustersizeTimeline}. 
In particular, we notice that TAMC allows to explore the under-critical as well as the post-critical domain of the nucleus size. 

A comment is in order about the results shown in Fig. \ref{fig:ClustersizeTimeline}, we notice that the system fluctuates between what appear to be metastable states in the post-critical domain. It might appear surprising that post-critical relatively stable nuclei do exist of a size smaller than the total number of atoms. 
{We propose two possible explanations of origin of this phenomenon. On the one hand, a further growth of clusters of effective nucleus size ${\mathcal N} = 400 - 600$, containing of the order of one half of the particles in the sample, requires a proper orientation of the nucleus with respect the simulation box otherwise the mismatch might prevent the formation of a perfect crystal containing all the particles in the sample. The re-orientation of a nucleus of this size is a slow process and therefore, in absence of any acceleration, do not take place over the time scale of a simulation. Another possible explanation is connected to the absence of additional collective variables controlling the global level of order in the growing cluster. The collective variable ${\mathcal N}(r_1, \dots, r_N)$ alone cannot control this phenomenon as it depends only on the value $C_{ij}$ of the nearest neighbor particles and therefore it controls only the local level of ordering. Moroni et al.\cite{PhysRevLett.94.235703} have tried to solve this problem by adding to their ${\mathcal N}(r_1, \dots, r_N)$  the collective variable ${\mathcal Q}_6^{cl} = 4 \pi / 13 \left ( \sum_{m=-6}^6 \left |{\mathcal Q}_{6m}^{cl} \right |^2 \right )^{1/2}$, where ${\mathcal Q}_{6m}^{cl}$ is the average of the local bond-orientational order parameter $q_{6m}(i)$ computed over the atoms belonging to the nucleus. However, Moroni et al., and Gasser et al.,\cite{science-292-258} who have performed experimental work on the nucleation of colloidal particles, have shown that the free energy barrier and the critical nucleus size can be accurately computed (measured) taking only into account the ${\mathcal N}(r_1, \dots, r_N)$ collective variable.}

We now turn to the reconstruction of the free energy curve vs. ${\mathcal N}$. Instead of running one long TAMC simulation we took advantage of parallel computers by running 32 independent TAMC trajectories. The 32 simulations were started from four configurations extracted from the TAMC trajectory shown in Fig. \ref{fig:ClustersizeTimeline}. Two of these four configurations correspond to under-critical nuclei (${\mathcal N} = 100$ and ${\mathcal N} = 200$) and two to post-critical nuclei (${\mathcal N} = 300$ and ${\mathcal N} = 400$). For each configuration we started eight TAMC trajectories with different initial random values of $\dot {z}$. These initial velocities were sampled from a Maxwell-Boltzmann distribution at ${\bar T}=30$ (the nucleation barrier as estimated by Moroni et al. at $T = 0.83$ is $\Delta F^* = 25 k_B T$). Each trajectory is evolved for 150000 collective variable steps.  In Fig. \ref{fig:FvsN} we report in logarithmic scale the histogram of $P({\mathcal N})$ ($-\log[P({\mathcal N})] \equiv \beta F({\mathcal N})$) vs. ${\mathcal N}$ computed along the TAMC trajectory. On the basis of this curve we can measure the nucleation free energy barrier and estimate the critical nucleus size. The free energy barrier we measure is $ \Delta F^* \sim 30.5 k_BT$, slightly higher than the free energy barrier computed by Moroni et al.\cite{PhysRevLett.94.235703} for a system at T=0.83 (corresponding to a 25\% degree of super-cooling)  by the Partial Path Transition Interface Sampling method\cite{moroni:4055} (PPTIS). 
It is intuitive to expect that to an higher super-cooling corresponds a lower free energy barrier, being the liquid state less stable in this condition. The decrease of the free energy barrier with the degree of super-cooling is indeed also a result of the classical theory of nucleation.\cite{KeltonGreer} This hypothesis is further supported by the fact that crystallization in highly supercooled systems can be studied by ``brute force'' simulations,\cite{PhysRevB.41.7042} which means that the free energy barrier, and therefore the corresponding timescale, decreases with the degree of supercooling.   

In spite of the difference in the free energy barrier, the critical size as estimated from our TAMC simulation is in good agreement with the one reported by Moroni et al.\cite{PhysRevLett.94.235703} The critical size, defined as the $\mathcal N$ corresponding to the maximum of the free energy, is $\sim 230$ in our simulations while it is 243 in 
Ref. [\citetext{15}]. 
Both, our and Moroni et al. results are strongly different from the critical size reported by ten Wolde et al.\cite{wolde:9932} (${\mathcal N}^* = 642$). However, in 
Ref. [\citetext{17}]  
the collective variable used for studying the nucleation is a different one (it is the ${\mathcal Q}_6$ bond-orientational order parameter of Steinhard et al.\cite{PhysRevB.28.784} of the entire sample) and the critical nucleus size is computed as the size of the maximum among the largest nuclei of crystal-like particles at the value of ${\mathcal Q}_6$  corresponding to the maximum of the free energy. 
\begin{figure}[h]
\begin{center}
\includegraphics[width=0.9\textwidth]{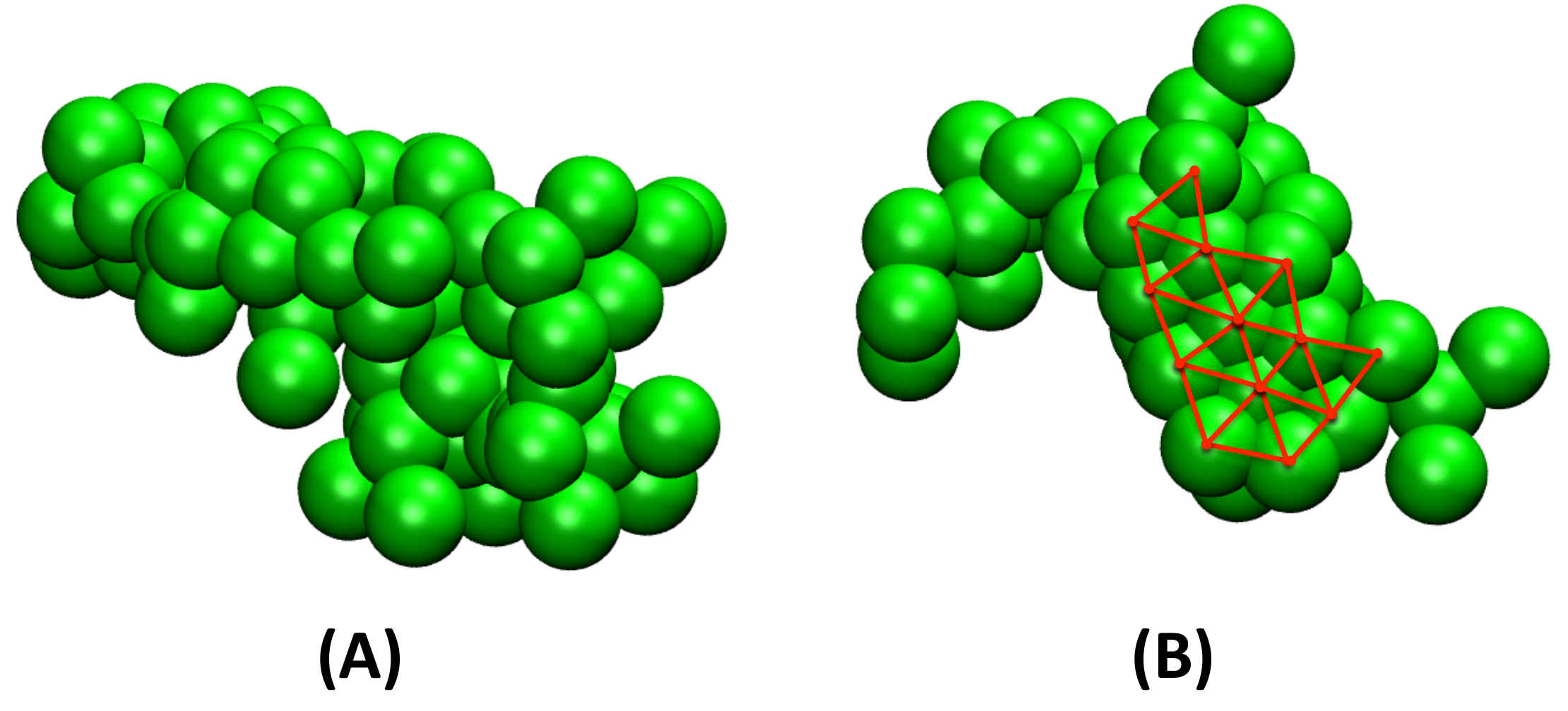}
\caption{ (Color online) Snapshots of an under-critical nucleus (A) and the ordered core of a post-critical nucleus (B). In panel (B) the network of bonds is superimposed to the atomistic structure to emphasize the ordered shape of the structure.}
\label{Fig:nucleiSnapshots}
\end{center}
\end{figure}
 
We also analyzed the structure of sub-critical and post-critical nuclei. We found that sub-critical (small) nuclei are, globally, disordered. An example of such a nucleus is shown in Fig. \ref{Fig:nucleiSnapshots}/A. This nucleus does not show any degree of global ordering. Rather, it looks liquid-like. This observation is partly in line with the results of 
Ref. [\citetext{15}] 
and Ref. [\citetext{17}], 
where it is reported that small nuclei are mainly liquid- and bcc-like. On the contrary, post-critical nuclei show a more ordered structure, with an fcc/hcp-like core.  In Fig. \ref{Fig:nucleiSnapshots}/B it is shown the ordered core of a post-critical nucleus, where the close-packed hexagonal crystal plane is clearly visible. This result is also consistent with recent experimental results\cite{science-292-258} on the nucleation of colloidal particles, which are characterized by post-critical nuclei with a rough surface and an ordered face centered cubic or hexagonal close-packed core.

Finally, we also run simulations at $P=5.6$ and $T = 1.2$. According to the phase diagram determined by Hansen and Verlet by MC simulations,\cite{PhysRev.184.151} at this pressure and temperature the most stable phase is the liquid phase. {We therefore expect the free energy curve to be minimal at low $\mathcal N$ and monotonically increasing with the nucleus size. Indeed, in our simulations in these condition the free energy is minimal at low $\mathcal N$, even though it seems to reach a plateau at large values. This is most likely due to the insufficient statistics we get for event of very low probability. However, our results confirm that at this pressure and temperature the most stable phase is the  liquid phase.}  

\begin{figure}[h]
\begin{center}
\includegraphics[width=0.45\textwidth]{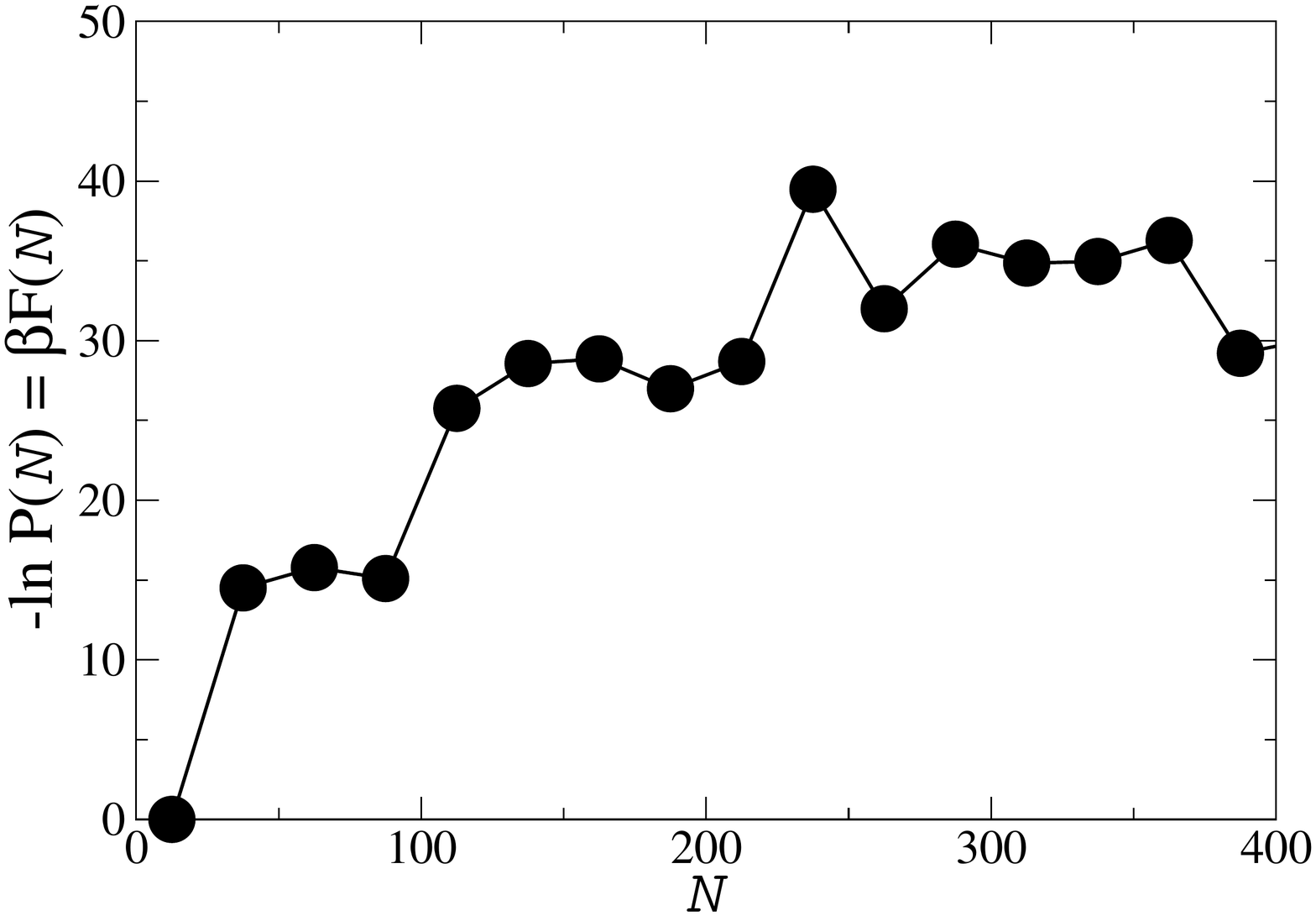}
\caption{Free energy vs. $\mathcal N$ curve at P=5.6 and T=1.2.}
\label{Fig:FvsN-liquid}
\end{center}
\end{figure} 
 
\section{Conclusions
\label{Conclusions}
}
In this paper we have introduced a new method for simulating rare events described by non-analytical collective variables of atomic positions. This kind of variables is intrinsic to {\em ab-initio} simulations, were the collective variables are the expectation value of the associated operators computed over the wavefunction corresponding to a given atomic configuration. However important cases of non-analytical variables appear also in classical simulations, as our illustration with the nucleation case has shown.

To illustrate the functioning of the method we studied the homogeneous crystallization in a sample of Lennard-Jones particles. The process has been studied using the new collective variable, Effective Nucleus Size $\mathcal N$ , introduced in this paper. Our results at $P=5.6$ and $T=0.92$ are in agreement with previous simulations. Moreover, simulations at a pressure and temperature in the liquid domain ($P=5.6$ and $T=1.2$) found a free energy curve growing with the nucleus size, in agreement with what we expect in the liquid domain. Our conclusion is that the method is ready for challenging applications. Work is in progress in this direction.

\section*{Acknowledgments}
The authors thank Eric Vanden-Eijnden for helpful discussions and suggestions. The authors wish to acknowledge SFI Grant 08-IN.1-I1869 which supported this work and the SFI/HEA Irish Centre for High-End Computing (ICHEC) for the provision of computational facilities.





\footnotesize{
\providecommand*{\mcitethebibliography}{\thebibliography}
\csname @ifundefined\endcsname{endmcitethebibliography}
{\let\endmcitethebibliography\endthebibliography}{}

}

\end{document}